\documentclass[runningheads]{llncs}
\usepackage[usenames,dvipsnames]{color} 
\usepackage{stfloats}
\usepackage{xcolor}
\usepackage[super]{nth}
\usepackage{mathtools}
\usepackage{wrapfig}
\usepackage{graphicx}
\usepackage{amsmath}
\usepackage{amssymb}
\usepackage{hyperref}
\usepackage{booktabs}
\usepackage{lipsum}

\usepackage[ruled,vlined]{algorithm2e}
\makeatletter
\newcommand{\printfnsymbol}[1]{%
  \textsuperscript{\@fnsymbol{#1}}%
}
\makeatother

\begin{document}

\title{Wakeword Detection under Distribution Shifts}
\titlerunning{Wakeword Detection under Distribution Shifts}

%\author{{Sree Hari Krishnan Parthasarathi\orcidID{0000-0002-5368-8860}, Lu Zeng\orcidID{0000-0002-3247-3844}, Christin Jose\orcidID{0000-0002-7594-4490}, Joseph Wang\orcidID{0000-0003-4571-6814}}}
\author{{Sree Hari Krishnan Parthasarathi, Lu Zeng, Christin Jose, Joseph Wang}}
\authorrunning{Parthasarathi, Zeng et al.}
\institute{Alexa, Amazon, USA\\
\email{\{sparta, luzeng, chrjse, wangjose\}@amazon.com}}

\maketitle

\begin{abstract}

We propose a novel approach for semi-supervised learning (SSL) designed to overcome distribution shifts between training and real-world data arising in the keyword spotting (KWS) task. Shifts from training data distribution are a key challenge for real-world KWS tasks: when a new model is deployed on device, the gating of the accepted data undergoes a shift in distribution, making the problem of timely updates via subsequent deployments hard. Despite the shift, we assume that the marginal distributions on labels do not change. We utilize a modified teacher/student training framework, where labeled training data is augmented with unlabeled data. Note that the teacher does not have access to the new distribution as well. To train effectively with a mix of human and teacher labeled data, we develop a teacher labeling strategy based on confidence heuristics to reduce entropy on the label distribution from the teacher model; the data is then sampled to match the marginal distribution on the labels. Large scale experimental results show that a convolutional neural network (CNN) trained on far-field audio, and evaluated on far-field audio drawn from a different distribution, obtains a 14.3\% relative improvement in false discovery rate (FDR) at equal false reject rate (FRR), while yielding a 5\% improvement in FDR under no distribution shift. Under a more severe distribution shift from far-field to near-field audio with a smaller fully connected network (FCN) our approach achieves a 52\% relative improvement in FDR at equal FRR, while yielding a 20\% relative improvement in FDR on the original distribution.
\end{abstract}

\begin{keywords} 
wakeword detection, distribution shifts, keyword spotting
\end{keywords}

\section{Introduction}
\label{sec:intro}
While deep learning models have a remarkable capacity to fit the training data distribution, fitting even random labels~\cite{Oriol}, distribution shifts from the training data can present challenges~\cite{Courville,Recht,Yossi}. These challenges include covariate shifts, prior shifts, selection bias, domain shift etc~\cite{Quionero-Candela}. Such distribution shifts can cause a model to learn spurious structures in the data and generalize poorly~\cite{Geirhos}. Our work in this paper happens in this context and focuses on small footprint on-device wakeword detection\footnote{Also known as keyword spotting; this is a task of detecting keywords of interest in a continuous audio stream.} models~\cite{Tucker,Fernndez,Jose,Panchapagesan,Sainath,Sun,Chen}. 

Wakeword detection is a modeling area that is prone to distribution shifts. For wakeword detection models deployed on device, the data that is accepted or rejected is gated by the model, causing a tight coupling between the model on device and the data available for subsequent training and evaluations. However, continuously selecting and labeling data can be time consuming and expensive. Beyond distribution shifts associated with deployed models, shifts can also occur due to changes in devices, user populations, and usage patterns/applications. An extreme version of this shift is the introduction of devices that differ significantly from existing population. Generalizing annotated data collected for one such set of devices to another requires addressing distribution shifts introduced. 

In this paper we investigate the problem of helping generalize an on-device wakeword detection model over a distribution shift caused by a model update and underlying temporal changes in data distributions. We propose a novel version of teacher/student training with two major deviations from the standard approach, where labeled training data is augmented with unlabeled data. First, since the teacher does not have access to the new distribution, our proposed labeling strategy reduces the entropy on the label distribution by using confidence heuristics, enabling student models to be trained from a combination of human and teacher labeled data. Second, assuming that the marginal distributions on labels do not change, we develop an approach to sampling unlabeled data to overcome both distribution shifts inherent to the KWS task as well as biases introduced by the proposed labeling scheme. 
To empirically validate the proposed approach we conducted two sets of large scale experiments (over 200K hours of unlabeled data) on de-identified production data. In these experiments, we demonstrate the ability to overcome both temporal distribution shifts as well as device-type distribution shifts.

\noindent\textbf{Related Work:} Research related to modeling under distribution shifts falls into three categories: a) without access to the shifted distribution, approached using robust optimization and meta-learning~\cite{Courville,arjovsky2019invariant}; b) with access to shifted distribution and a small set of labeled data, approached using active learning~\cite{chelba2006adaptation}; c) with access to the shifted distribution, but without labels - our problem falls under the third category. Continual learning~\cite{parisi2019continual} is a related topic where the data is constrained to be processed online. 

Learning from unlabeled data, especially under distribution or domain shifts, remains a challenge~\cite{ruder2018strong,elsahar2019annotate}. Zhao et al~\cite{zhao2020robust} imposed similarity constraints on labeled and unlabeled distributions as a form of \textit{consistency regularization} principle, proposed in~\cite{bachman2014learning,sajjadi2016regularization,Berthelot}. Our algorithm falls under this category, imposing the assumption that the marginal distributions on labels do not change; however the previous approaches perturb the data in an unbiased fashion, whereas in our setting, the data distribution shifts. Another principle used in SSL, \textit{entropy minimization}, encourages the pseudo-labels to be well-separated on the unlabeled data, through the use of regularizers~\cite{grandvalet2004semi,lee2013pseudo}. Inspired by this principle, but to scale it to a large unlabeled dataset, our algorithm uses confidence thresholds to encourage separability of classes. Our approach combines these two principles to handle distribution shifts in a teacher/student framework where the teacher has not seen the new distribution too.

\section{Distribution Shifts in Wakeword Data}
\label{sec:distribution_shifts}
\begin{figure}[t]
  \centering
  \includegraphics[width=\linewidth]{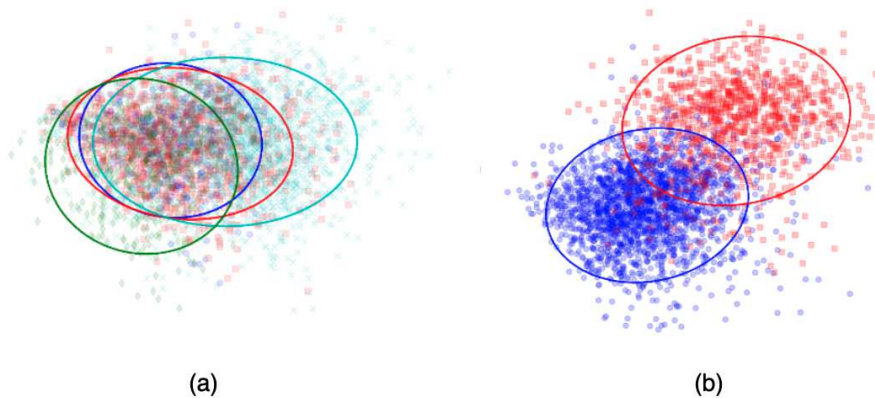}
  \caption{\textit{(a) Embeddings of wakeword data plotted with respect to year along with the first standard deviation contour of a Normal distribution approximation. Data from 2018 is represented by blue, from 2019 by red, 2020 by cyan, and 2021 by green; (b) Embeddings of random samples of wakeword data drawn from a far-field device (blue) and mobile devices (red) along with the first standard deviation contour of a Normal distribution approximation.}}
  \label{fig:temporal_device_distribution_shift}
\end{figure}

We illustrate two shifts in data distributions below. First, Fig.\ref{fig:temporal_device_distribution_shift} (a) shows the temporal drift in the distribution of data the model observes due to factors such as the wakeword model gating behavior (as described in the introduction), physical changes to devices, introduction of newer versions of devices with differing hardware/software, growth and change in user population, and change in use of devices. This introduces the challenge of leveraging older annotated data that may not be representative of the current distribution.

A second major shift in wakeword data distributions occurs due to generalization to differing devices/use cases. An example of this is shown in Fig. \ref{fig:temporal_device_distribution_shift}(b), where data collected from mobile devices is compared to data collected on far-field devices. Mobile device data is fundamentally different for a wide range of factors. Differing customer usage leads to different acoustic and background noise conditions (e.g. use in automobiles, while walking around), generally is near-field, and has significantly differing microphones and audio front-end capabilities. In this setting, our goal is to build performant models for new device-types with limited annotated data by leveraging annotated data from different device-types.

\section{On-device Wakeword Detection Models}
\label{sec:model_architectures}
We describe the two wakeword detection models, namely the CNN and FCN. Models are similar to~\cite{Sainath,Chen}, with the difference being that the input window encompasses a larger audio context. The models are trained using a set of positive and negative examples (i.e. positive examples contain the wakeword, while the negatives do not). During training, the wakeword is consistently center aligned in the input window~\cite{Jose}. 

\begin{figure}[t]
  \centering
  \includegraphics[width=0.85\linewidth]{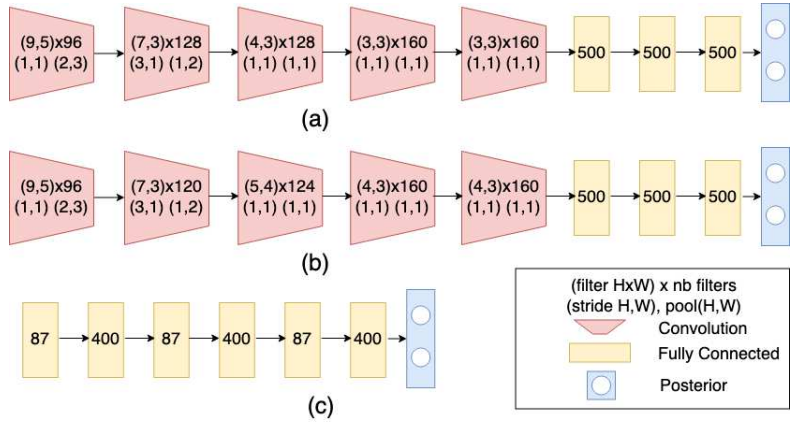}
  \caption{\textit{Model architectures: (a) CNN annotation model with 5 CNN layers and 3 fully-connected layers (4M parameters); (b) CNN on-device wakeword detection model with 5 CNN layers and 3 fully-connected layers (2M parameters); (c) FCN on-device wakeword detection model with 5 fully-connected layers (250k parameters).}}
  \label{fig:model_architecture}
\end{figure}

The CNN model architecture has 2M learnable parameters, and it operates on 64-dimensional log mel filter bank energy (LFBE) features, computed with analysis window size and shift of 25 ms and 10 ms respectively; the input to the model is 100 frames. The model has 5 CNN layers and 3 fully-connected layers, as shown in Fig. \ref{fig:model_architecture} (b). Dropout and batch normalization are used in all the layers. The output is a binary classification layer trained with cross entropy loss, representing the probability of ``wakeword'' and ``non-wakeword''~\cite{gao2020front}.

The FCN model architecture has 250K learnable parameters, and it operates on 20-dimensional LFBE features computed over the same analysis window size and shift as the CNN; the input to the model is 81 frames, downsampled by a factor of 3. The architecture consists of five fully connected layers. Details of this model architecture is presented in Fig. \ref{fig:model_architecture} (c). Dropout and batch normalization are used with all hidden layers. Similar to CNN, the output is a binary classification layer trained with cross entropy loss.

During inference, the posterior estimates, from CNN or FCN, corresponding to the wakeword are smoothed by an exponential moving average (EMA) or a windowed smoothing average (WMA) filter respectively; these are then thresholded to infer the wakeword hypothesis.

\section{Proposed SSL Algorithm}
\label{sec:sys_description}
The SSL algorithm consists of selection of an annotation model, selection of thresholds to compute pseudo-labels, distributing matching with subsampling, and the actual training. We describe each of these components in depth.

\noindent\textbf{Selecting and Annotating Data:} A critical aspect of model training to account for distribution shifts in the selection of unlabeled training data to be annotated with a teacher model. Let $z^o_s = (x^o_s, y^o_s)$ be the original supervised training data distribution. This supervised data is not drawn from the distribution of examples the model is exposed to due to both access to the data, as well as the imbalance of classes in the detection setting. Instead, the supervised data selection is more heavily drawn from ``hard'' examples that are close to the decision boundary of previous models. Note that an example can be ``hard'' due to a number of reasons beyond distribution shift; for example, features not being able to separate examples.

Under a distribution shift, we ideally would like to have access to $z^n_s = (x^n_s, y^n_s)$, the equivalent supervised training data distribution drawn heavily from ``hard'' examples from the new shifted distribution, with subsequent models then trained using  $\{z^o_s \cup z^n_s\}$. Since manual labeling of hard examples is slow, we want some approximation of $z^n_s$ using an SSL algorithm. Let $(z^n_u, s^n_u) = (x^n_u, y^n_u, s^n_u)$ be the unsupervised distribution with score $s^n_u$ from a teacher model. We want to approximate $z^n_s$ with $z^n_u$. We make the assumption that the marginal distributions on labels (i.e. $y^o_s$ and $y^n_s$) share the same distributions, even under distribution shifts.

In the absence of distribution shifts, we could use a teacher model to derive soft or hard pseudo labels. However in the presence of distribution shifts, selecting data that is hard for the student will yield noisy labels since teacher model has not been trained on the shifted distribution. To ameliorate this issue, we design our algorithm so that the teacher is able to label well, applying thresholds $\tau_+$ and $\tau_-$ to obtain $z^n_{u_{\tau}} := (x^n_{u_{\tau}}, y^n_{u_{\tau}}) = f_\tau(x^n_u, y^n_u, s^n_u)$.

The thresholded scores on unlabeled examples from the annotation model are binarized to pseudo-labels: data above the accept threshold ($\tau_{+}$) have positive labels, while those below the reject threshold ($\tau_{-}$) have negative labels. Unlabeled data with scores between the two thresholds are then discarded. 

The accept/reject thresholds are determined based on analysis done on held-out examples that have groundtruth. We investigated precision and false positive rate (FPR) as means to quantify the ``purity'' of positive labels. Since data is biased towards positive examples, precision does not capture purity with sufficient granularity: therefore, we choose an accept threshold value on FPR. Similarly, FRR and False Omission Rate (FOR) were considered as metrics for negative labels. Analysis on held-out data showed that both FOR and FRR are reasonable metrics; we choose a reject threshold value on FRR.

Applying the thresholds $\tau_+$ and $\tau_-$ can be viewed as a form of entropy minimization, however this strategy alone is not sufficient to approximate $z^n_s$ with $z^n_{u_{\tau}}$. Since we have access to $y^o_s$ (and we assume $y^o_s$ and $y^n_s$ share the same distribution), we subsample $z^n_{u_{\tau}}$ to yield $z^n_{u_{\tau, \theta}} := (x^n_{u_{\tau, \theta}}, y^n_{u_{\tau, \theta}}) = g_\theta(f_\tau(x^n_u, y^n_u, s^n_u))$. Subsampling of the data can be viewed as a form of consistency regularization, where the distributions of $y^o_s$ and $y^n_s$ are kept close through the selection of data.

\begin{algorithm}[t]
     \KwData{$\tau_{+}$, accept threshold\newline
            $\tau_{-}$, reject threshold\newline
            $\theta_{+}$, positive class subsampling\newline
            $s \in \mathcal S $, unlabeled data\newline
            $score$, annotation model posterior}
 \KwResult{Pseudo-labeled data}
 initialization\;
\For {$s \in \mathcal S $} {
  score = query\_score(s)\;
  u = rand()\;
  \uIf{score $\geq \tau_{+}$ \text{and} u $\leq \theta_{+}$}{
   label = ``Wakeword''\;
   }\uElseIf{score $\leq \tau_{-}$}{
   label = ``Not Wakeword''\;
  }\Else{return None\;}
 }
 \caption{SSL scheme to generate pseudo-labels on unlabeled data}
 \label{alg:label_ssl}
\end{algorithm}

\noindent\textbf{Training Method:} We use the verification model described in Fig. \ref{fig:model_architecture} (a) as the annotation model: it has a similar architecture to the CNN model described earlier, with 64-dimensional LFBE feature input to 5 stacking CNN layers fed to 3 fully-connected layers; the last layer performs the binary classification task, however the input window context is larger (consisting of 195 frames) and the model is specifically trained for the verification task rather than general wakeword detection. The model parameters are optimized with the cross entropy loss, and the model outputs a posterior probability~\cite{Kumar}. The posteriors from the annotation model are used to annotate unlabeled data with hard labels; we refer to these as scores.

Similar to~\cite{one_million_hr,Berthelot}, we interleave this teacher annotated data with human labeled data within a minibatch during training. We set a weight of $\lambda:(1-\lambda)$ for labeled and pseudo-labeled data respectively.

\section{Experimental Setup}
\label{sec:experimetal_setup}
All experiments in this paper were conducted on de-identified production datasets. We now describe the setup for the two sets of training and evaluation experiments: a) CNN trained on far-field audio, and evaluated on far-field audio with and without distribution shift; b) FCN trained on far-field audio, and evaluated on near and far-field audio.  We also discuss the evaluation metric used in our experiments. For all datasets, pseudo-labels are generated using annotation models as described in Section \ref{sec:sys_description}; in the interest of space, we do not discuss these annotation models as the training has been described in~\cite{Kumar}.

\noindent\textbf{Datasets:} For the two sets of experiments, we created a labeled and a pseudo-labeled training dataset; the latter was constructed using the procedure described in Section~\ref{sec:sys_description}. For CNN: a) the labeled data consisted of 8K hours of far-field audio; b) the pseudo-labeled data consisted of 200K hours of far-field audio. Note these datasets are drawn from different distributions, as discussed earlier. For FCN: a) the labeled data consisted of 12K hours of far-field audio; b) the pseudo-labeled data consisted of 16K hours of near-field mobile phone data.

We created evaluation datasets with and without distribution shifts for the two cases. For CNN: a) without distribution shift labeled data consisting of 3K hours of far-field audio drawn from the same distribution as the labeled training data; b) distribution shifted labeled data consisting of 1K hours of far-field audio drawn from the same distribution as the pseudo-labeled training data. For FCN: a) without distribution shift labeled data consisting of 33 hours of far-field audio drawn from the same distribution as the labeled training data; b) distribution shifted labeled data consisting of 30 hours of near-field mobile phone audio drawn from the same distribution as the pseudo-labeled training data.

\noindent\textbf{Training and Evaluation:} We select the accept/reject thresholds and the subsampling factor for the positive class for the two sets of model training based on held-out datasets. We tuned $\lambda$, the mixing factor, for CNN and FCN training, obtaining a smaller $\lambda$ for FCN training (reasonable given the extent of distribution shift). During model training, with each minibatch, we updated the model using the Adam optimizer to compute the error signal; 700K model updates were done for CNN, while 200K model updates were done for FCN. During inference, we tuned the EMA and WMA values for the models on held-out datasets. 

We use DET curves to measure the performance of the models, using False Rejection Rate (FRR) and False Discovery Rate (FDR); DET curves for proposed models are plotted compared to their baselines. Similar to ~\cite{Sun,gao2020front,Tucker,Jose,Kumar}, we normalize the axes of the DET curves using the baseline model's operating point (OP). For certain results, in the interest of space, we only report the results in terms of relative FDR at the baseline model's OP.

\section{Results}
We now describe the results for two sets of experiments presented in Section~\ref{sec:experimetal_setup}: a) far-field to far-field distribution shifts with CNN; b) far-field to near-field distribution shifts with FCN. We discuss the first experiment in depth using an ablation study with respect to the size of pseudo-labeled datasets as well as an analysis using the reliability of annotations.
\label{sec:results1}

\subsection{Far-field Distribution Shifts with CNN}
The baseline and the proposed approach use the CNN model architecture discussed in Section~\ref{sec:model_architectures}, with results presented in Table~\ref{tab:CNN_rst}. While the baseline model was trained on labeled data, the proposed model was trained on a mixture of labeled and pseudo-labeled data. As can be seen from the table, the proposed approach achieves a 14.3$\%$ relative improvement in FDR at FRR matching the baseline model on eval data with a distribution shift (matching the distribution of the pseudo-labeled data). Furthermore, it achieves a 5.0$\%$ relative improvement in FDR on eval data without a distribution shift (matching the distribution of the labeled data).

\begin{table}
  \caption{\textit{Rel. imp. in FDR at FRR matching the baseline model on eval data with and without distribution shifts. Baseline model was trained on only labeled data while the model using proposed approach was trained on both labeled and pseudo-labeled data.}}
  \label{tab:CNN_rst}
  \centering
 \begin{tabular}{lc}
    \multicolumn{1}{c}{\textbf{\shortstack{Condition}}} &  \multicolumn{1}{c}{\textbf{\shortstack{Rel. FDR Imp.  (\%)}}}   \\
    \hline
    Distribution shift        &    14.3           \\
    \hline
    No distribution shift          &   5.0          \\
    \hline
  \end{tabular}
\end{table}

\begin{figure}[t]
  \centering
  \includegraphics[width=\linewidth]{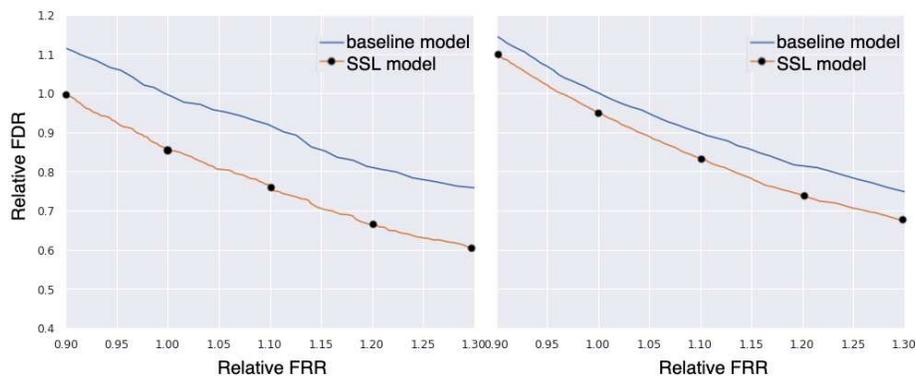}
  \caption{\textit{DET curves showing the baseline (blue) and the SSL (orange/with dots) models evaluated on a distribution shift (left) and no distribution shift (right).}}
  \label{fig:cnn_ds}
\end{figure}

We also present the full DET curves for both the baseline and the proposed model in Fig.~\ref{fig:cnn_ds}. Note that $(1.0, 1.0)$ on the DET curves correspond to the OP of the baseline model. It can be seen from Fig.~\ref{fig:cnn_ds} that the DET curves mirror the larger and smaller gains, respectively, on evaluation data sets observed in the table; these correspond to conditions with and without distribution shifts. This shows that the proposed approach generalizes well.

\begin{figure}[t]
  \centering
  \includegraphics[width=\linewidth]{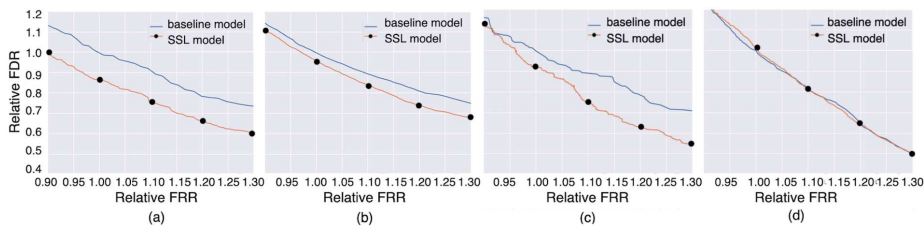}
  \caption{\textit{Baseline (blue) and the SSL (orange/with dots) models evaluated on: (a) easier examples with distribution shift; (b) easier examples with no distribution shift; (c) harder examples with distribution shift; (d) harder examples with no distribution shift.}}
  \label{fig:cnn_easy_and_hard}
\end{figure}

\begin{figure}[t]
  \centering
  \includegraphics[width=\linewidth]{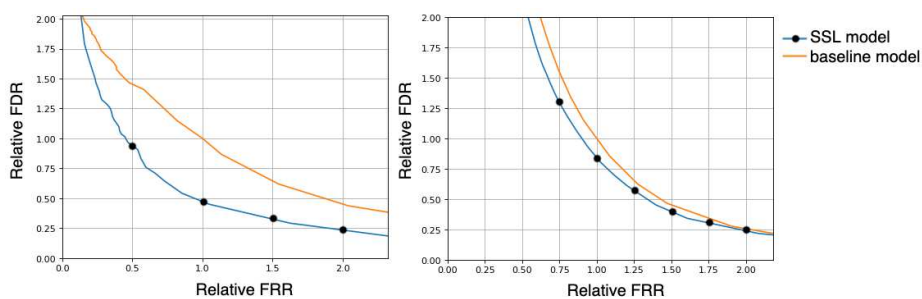}
  \caption{\textit{Comparing the baseline (orange) and the SSL (blue/with dots) models evaluated on a distribution shift (left) and no distribution shift (right).}}
  \label{fig:250k_dnn_plot}
\end{figure}

\subsubsection{Effect of Pseudo-Labeled Training Data Size}

\begin{table}
\caption{\textit{Rel. imp. in FDR at FRR matching the baseline model on eval data with and without distribution shifts. The size of the pseudo-labeled training data is increased.}}
    \label{tab:learning_curve}
 \centering
 \begin{tabular}{ccc}
  \multicolumn{1}{c}{\textbf{\shortstack{Data (hrs)}}} & \multicolumn{1}{c}{\textbf{\shortstack{Relative FDR \\(Distr. Shift)}}} & \multicolumn{1}{c}{\textbf{\shortstack{Relative FDR \\(No Distr. Shift)}}}  \\
  \hline
1K   & $-16.8$ & $-22.5$ \\
 \hline
 4K   & $-1.8$ & $-3.2$ \\
 \hline
 16K   & $4.7$ & $0.7$ \\
 \hline
 64K   & $10.1$ & $1.5$ \\
 \hline
 200K & $9.6$ & $1.3$ \\
  \hline
 \end{tabular}
\end{table}

Given the small model size, we wanted to understand how much pseudo-labeled data was useful. Table~\ref{tab:learning_curve} presents performance of the CNN models in terms of increasing sizes of the pseudo-labeled data. To avoid the impact of imbalanced training set sizes, no labeled data was used. Results are reported on evaluation data (both with and without a distribution shift) in terms of relative improvement in FDR at equal FRR compared to the baseline model (from Table~\ref{tab:CNN_rst}). We see a steady improvement in performance on both eval data sets till the training data increases to 64K hours; thereafter the performance saturates indicating the capacity of the model.

\subsubsection{Effect of Subsampling on Pseudo-Labeled Training Data}

\begin{table}
    \caption{\textit{Rel. imp. in FDR at FRR matching the baseline model on eval data with and without distribution shifts for models trained without (-) and with (+) subsampling.}}
  \label{tab:1M_ssl_exp}
  \centering
 \begin{tabular}{lcc}
    \multicolumn{1}{c}{\textbf{\shortstack{Models}}} & \multicolumn{1}{c}{\textbf{\shortstack{Rel. FDR Imp ($\%$) \\(Dist. Shift)}}} & \multicolumn{1}{c}{\textbf{\shortstack{Rel. FDR Imp ($\%$)\\(No Dist. Shift)}}} \\
    \hline
    \multicolumn{1}{l}{\shortstack{$-$ subsampling}}      &   $2.0$ & $-21.0$  \\
    \hline
    \multicolumn{1}{l}{\shortstack{$+$ subsampling}}      & $4.0$ & $-3.0$    \\ 
   \hline
  \end{tabular}
\end{table}

For studying the effect of subsampling, we trained a baseline model only on 1K hours of labeled audio. Two models were trained only on 1K hours of pseudo-labeled audio using the approach in Section~\ref{sec:sys_description}: a) a model that uses the best subsampling factor identified in Section~\ref{sec:experimetal_setup}; b) a model that uses no subsampling. Table~\ref{tab:1M_ssl_exp} presents the results in terms of relative improvement in FDR at equal FRR compared to the baseline model. With distribution shift, the pseudo-labeled models are better, independent of the subsampling factor; not surprisingly, on the original distribution, the models trained on pseudo-labeled data are worse. However subsampling yields a large improvement on the original distribution and bridges the gap with the baseline.

\subsubsection{Effect of Pseudo-Labeling on Easy and Hard Cases}
A risk of using pseudo-labeled examples the annotation model was confident on is whether the model performance generalized well over ``hard'' cases. We divided the eval data from both the shifted and unshifted distributions based on whether it was hard or easy for the annotator to label. Fig.~\ref{fig:cnn_easy_and_hard} presents the DET curves for those cases, with and without distribution shifts.

Fig.~\ref{fig:cnn_easy_and_hard} (a, b) shows that for easier examples the proposed method yields improvements with and without distribution shifts. We see the same trend as in Fig.~\ref{fig:cnn_ds}, where the gains were larger on the setting with distribution shift. For harder examples (Fig.~\ref{fig:cnn_easy_and_hard} (c, d)), the trend is similar, except that we observe no gains when there is no distribution shift, demonstrating the proposed method generalizes to ``hard'' examples.

\label{sec:results2}
\begin{table} [t]
 \caption{\textit{Rel. imp. in FDR at FRR matching the baseline model on eval data with and without distribution shifts.}}
    \label{tab:FCN_rst}
  \centering
 \begin{tabular}{lc}
    \multicolumn{1}{c}{\textbf{\shortstack{Condition}}} & \multicolumn{1}{c}{\textbf{\shortstack{Rel. FDR Imp. (\%)}}}   \\
    \hline
    Distribution shift         &  $52.0$           \\
    \hline
    No distribution shift          &  $20.0$           \\
   \hline
  \end{tabular}
\end{table}

\subsection{Far and Near-Field Distribution Shifts with FCN}

For the second set of experiments, the baseline and the proposed approach use the FCN model architecture discussed in Section~\ref{sec:model_architectures}. Table~\ref{tab:FCN_rst} presents the results. The proposed approach achieves a 52$\%$ relative improvement in FDR at FRR matching the baseline model on eval data with a distribution shift. Furthermore, it yields a $20.0\%$ relative improvement in FDR without a distribution shift, showing that the proposed approach generalizes well even under a severe distribution shift even with models having much lower capacity. The gains observed in the table also reflects on Fig.~\ref{fig:250k_dnn_plot} with DET curves corresponding to conditions with and without distribution shifts.

\section{Conclusions}
\label{sec:conclusions}
This paper characterizes distribution shifts in wakeword detection and proposes an approach to address it. Utilizing large scale unlabeled data from the shifted distribution, we use an annotation model with guidance from accept/reject confidence heuristics to generate pseudo-labels. We mitigate the over representation of subsets of data on which the annotation model does well by subsampling the positive class conditional distribution. Experiments on de-identified production data show that for a CNN model (2M parameters) trained on far-field audio and evaluated on far-field audio drawn from a different distribution, the proposed approach achieves a 14.3\% relative improvement in FDR at equal FRR, while still yielding a 5\% improvement in FDR under no distribution shift. We performed 3 ablation studies: a) size of pseudo-labeled data; b) with and without subsampling; c) easy and hard cases, confirming the viability of the proposed approach. As a second study, under a more severe distribution shift from far-field to near-field audio, with a smaller footprint FCN (250K parameters), our approach achieves a 52\% relative improvement in FDR at equal FRR, while yielding a $20\%$ relative improvement in FDR on the original far-field distribution.

\iffalse
%\small
\section{Acknowledgements}
We would like to thank Dilek Hakkani-Tur, Aishwarya Padmakumar, Brian Kulis, Noah Stein, Rajath Kumar, Peng Liu, Reza Solgi, Harish Mallidi, Penny Karanasou, and Sri Karlapati for feedback on the manuscript.
\fi

%References on new page

\bibliographystyle{IEEEtran}
\bibliography{tsd1093a}

\end{document}